\documentstyle[11pt,epsfig]{article}
\title{\bf Effect of GUP on the Kepler problem and a variable minimal length}
\author{Fatemeh Ahmadi $^{1}$\thanks{email:
fahmadi@srttu.edu} and Jafar Khodagholizadeh $^{2}$\thanks{email:
j.gholizadeh@modares.ac.ir}. \\$^1${\small Department of Physics, Shahid
Rajaee Teacher Training University, Lavizan, Tehran 16788, Iran}\\$^2${\small Department of Physics, Tarbiat Modares University
P.O.Box:14115-398 Tehran, Iran}\\}

\begin{document}

\maketitle

\begin{abstract}
Various approaches to quantum gravity, such as string theory, predict a minimal measurable length and a modification of the Heisenberg Uncertainty Principle near the Plank scale, known as the Generalized Uncertainty Principle (GUP). Here we study the effects of GUP which preserves the rotational symmetry of the spacetime, on the Kepler problem. By comparing the value of the perihelion shift of the planet Mercury in Schwarzschild-de Sitter spacetime with the resulted value of GUP, we find a relation between the minimal measurable length and the cosmological constant of the spacetime. Now, if the cosmological constant varies with time, we have a variable minimal length in the spacetime. Finally, we investigate the effects of GUP on the stability of circular orbits.\\


\end{abstract}
\vspace{.7cm}

\section{Introduction}
An important consequence of quantum gravity scenarios is the existence of a minimum length. Since the Heisenberg uncertainty principle
does not exert any restriction on the measurement precision of the particle's positions or momenta,  there is no minimum measurable length
in the usual Heisenberg picture. In the past few years, quantum mechanics with modification of the usual canonical commutation relations has been
investigated intensively\cite{Amati}-\cite{Pedram}. Such works, which are motivated by string theory and quantum gravity, suggest a minimal length as $\Delta x\geq\hbar\sqrt{\beta}$. This means that there is no possibility to measure coordinate $x$ with accuracy smaller than $\hbar\sqrt{\beta}$, where $\beta$ is a positive parameter and depends on the expectation value of the position and momentum operators \cite{Garay, Kempf}. Since in string theory the minimum observable distance is the string length, $\sqrt{\beta}$ is proportional to this length. If we set $\beta=0$, the usual Heisenberg algebra is recovered. In low energy limit, these quantum gravity effects can be neglected, but in circumstances such as the very early universe or in the strong gravitational field of a black hole one has to consider these effects. The modification induced by the generalized uncertainty principle (GUP) on the classical orbits of particles in a central force potential has been considered in \cite{Benczik}.

Alternatively, it  could be argued that a fundamental length in nature may have dramatic impacts on our universe. Also, it could be seen that this length may be related to vacuum energy, i.e. dark energy \cite{Calmet}. The introduction of $\Lambda$ in the formalism of GUP has been studied in the literature \cite{Chang}-\cite{AKempf}. Most of different analyses are performed by introducing an ultraviolet (UV) cut-off on the planck scale\cite{Chang}. But, other works introduce an infrared (IR) cut-off by using the so-called q-Bargmann Fock formalism \cite{Che, Adler, AKempf}. In this formalism, if we impose the $l_{pl}$ as UV cut-off and the cosmological constant $\Lambda$ as an IR one, the UV scale is dual to the IR one.

In this paper, we are going to proceed to study the effects of GUP on the Kepler problem. The main consequence of this work is the constraint imposed on the minimal observable length and the parameter $\beta$ in comparison with observational data of Mercury. Also, by comparing the value of the perihelion shift of Mercury in Schwarzschild-de Sitter spacetime with the resulted value of GUP, we obtain a relation between the minimal length and the cosmological constant. Now if the cosmological constant is time dependent, we will have a minimal length to vary with time. This matter could be important in quantum gravity scenarios for studying the early universe and black hole physics. Finally, we investigate the stability of circular orbits of planets in this framework.  We show when the angular momentum of the planet is large, the condition for stability of circular orbits differs considerably from classical mechanics.
\section{The modified Hamiltonian equation }

We consider the commutation relation with the GUP as
\begin{equation}
[x_{i},p_{j}] = i \hbar (1+\beta p^{2})\delta_{ij},\label{eq1}
\end{equation}
this commutation relation implies the minimal length uncertainty relation
\begin{equation}
\Delta x \geq \frac{\hbar}{2} (\frac{1}{\Delta p}+\beta \Delta p),\label{eq2}
\end{equation}
which appears in perturbative string theory \cite{Garay, Kempf}. If the components of the momentum  $  p_{i} $ are assumed to commute with each other
\begin{equation}
[p_{i},p_{j}]=0 ,
\label{eq3}
\end{equation}
then the commutation relation among the coordinate  are determined as
\begin{equation}
[x_{i},x_{j}]=2 i\hbar\beta (p_{i}x_{j}-p_{j}x_{i}).
\label{eq4}
\end{equation}
Since the commutation relations (\ref{eq1}), (\ref{eq3}) and (\ref{eq4}) do not break the rotational symmetry, the generators of rotations can be expressed
in terms of the position and momentum operators as follows
\begin{equation}
L_{ij}\equiv\frac{1}{1+ \beta p^{2}}(x_{i}p_{j}-x_{j}p_{i}).\label{eq10}
\end{equation}
In the  classical limit, using the Poisson bracket, we can write equations (\ref{eq1}), (\ref{eq3}) and (\ref{eq4}) as follow
\begin{eqnarray}
\{ x_{i},p_{j}\} &=& (1+\beta p^{2})\delta_{ij} , \nonumber\\
\{ p_{i},p_{j}\} &=& 0 , \nonumber\\
\{ x_{i},x_{j}\} &=& 2\beta(p_{i}x_{j}-p_{j}x_{i}).\label{eq5}
\end{eqnarray}

The Poison bracket must possess the same properties as the quantum mechanics commutators, namely, it must be anti-symmetric, bilinear and satisfy the Leibniz rules and the Jacobi Identity. Therefore, we can derive the general form of the Poisson bracket for any functions of the coordinates and momenta as
\begin{eqnarray}
\{F, G\}&=& (\frac{\partial F}{\partial x_{i}}\frac{\partial G}{\partial p_{j}}-\frac{\partial F}{\partial p_{i}}\frac{\partial G}{\partial x_{j}}) \{x_{i},p_{j}\} +
\frac{\partial F}{\partial x_{i}}\frac{\partial G}{\partial x_{j}}\{x_{i}, x_{j}\},\label{eq6}
\end{eqnarray}
also, we can obtain the time evolutions of the coordinates and momenta as follows
\begin{eqnarray}
\dot{x_{i}}&=&\{x_{i}, H\}=\{x_{i}, p_{j}\}\frac{\partial H}{\partial p_{j}}+ \{x_{i}, x_{j}\}\frac{\partial H}{\partial x_{j}}, \nonumber\\
\dot{p_{i}}&=&\{p_{i}, H\}=-\{x_{i}, p_{j}\}\frac{\partial H}{\partial x_{j}}.\label{eq7}
\end{eqnarray}
Now we can use these equations to study the motion of a macroscopic object. In this paper, we consider Hamiltonian of a particle in a central potential
\begin{equation}
H=\frac{p_{i}^{2}}{2m}+ V(r),\label{eq8}
\end{equation}
the time evolution of the coordinates and momentums became
\begin{eqnarray}
\dot{x_{i}}&=&(1+\beta p^{2})\frac{p_{i}}{m}-2\beta(\frac{1}{r} \frac{\partial V}{\partial r})L_{ij}x_{j} , \nonumber\\
\dot{p_{i}}&=&-(1+\beta p^{2})x_{i} (\frac{1}{r} \frac{\partial V}{\partial r}),
\label{eq9}
\end{eqnarray}
where we have neglected  terms of the order of $\beta^{2}$. For motion in central force potential, the $ L_{ij}$'s  are conserved due to rotational symmetry $ \{ L_{ij},H \}=0 $. The conservation of the $L_{ij}$'s imply that the motion of the particle will be confined to a 2-dimensional plane spanned by the coordinate and momentum vectors at any point in time. Now, by using  equations (\ref{eq9}) we can drive the equation of motion of a particle
\begin{equation}
m \ddot{x_{i}}=(1+3\beta p^{2}) \dot{p_{i}}-2\beta (\frac{1}{r}\frac{dV}{dr})p_{i}.\label{eq11}
\label{09}
\end{equation}
In the next section, we  study the Kepler problem in this framework.
\section{GUP and Kepler problem }
As we know that the Kepler potential is  $ V(r)=-\frac{K}{r} $, so the equation (\ref{eq11}) became
\begin{equation}
m \ddot{x_{i}}=-(1+4\beta p^{2})\frac{x_{i}}{r}\frac{K}{r^{2}}-2\beta(\frac{K}{r^{3}})p_{i},
\label{eq12}
\end{equation}
since $ \beta $ is a very small parameter and $ r $ is very large, we can ignore the second term in comparison with the first term. In spherical coordinates, the equations above can be written as
\begin{equation}
m(\ddot{r}-r\dot{\theta}^{2}-r\dot{\varphi}^{2} \sin^{2}\theta)=-{(1+ 4 \beta m^{2} (\dot{r}^{2}+r\dot{\theta}^{2}+ r^{2}\dot{\varphi}^{2} \sin^{2}\theta))}\frac{\partial V}{\partial r},
\label{eq13}
\end{equation}
\begin{equation}
m\frac{d}{dt}(r^{2} \dot{\theta})-m r^{2} \dot{\varphi}^{2} \sin\theta \cos\theta \dot{\varphi}^{2}=0,
\label{eq14}
\end{equation}
\begin{equation}
\frac{d}{dt}(m r^{2}\dot{\varphi} \sin^{2}\theta)=0.
\label{eq15}
\end{equation}
We assume the case of the plane orbit that is a valid solution, and for simplicity we consider the equatorial orbits i.e $ \theta=\frac{\pi}{2} $. Then the above equations become as follows
\begin{equation}
m(\ddot{r}-r \dot{\varphi}^{2})= - \frac{K}{r^{2}}- 4 \beta m^{2} ( \dot{r}^{2}+r^{2} \dot{\varphi}^{2} )\frac{K}{r^{2}},
\label{eq16}
\end{equation}
\begin{equation}
m \frac{d}{dt}(r^{2} \dot{\theta})=0,\label{eq17}
\end{equation}
\begin{equation}
\frac{d}{dt}(m r^{2}\dot{\varphi})=0.\label{eq18}
\label{7}
\end{equation}
According to equation (\ref{eq18}), $ L=mr^{2}\dot{\varphi} $ is a constant of motion so  equation (\ref{eq16}) becomes
\begin{equation}
m \ddot{r}=\frac{L^{2}}{m r^{3}}( 1- 4 \beta m \frac{K}{r})-\frac{K}{r^{2}}- 4 \beta K m^{2} (\frac{\dot{r}^{2}}{r^{2}}).
\label{eq19}
\end{equation}
Now, for the right hand side of the equation above, we consider a new potential as
\begin{eqnarray}
V_{new}(r)&=&\frac{L^{2}}{2mr^{2}}-\frac{ 4 K \beta}{3}\frac{L^{2}}{r^{3}}-\frac{K}{r}- 4 \beta K m^{2}\frac{\dot{r}^{2}}{r}\nonumber\\
 &=& \frac{L^{2}}{2mr^{2}}(1-\frac{8m\beta}{3}\frac{K}{r})-\frac{K}{r}- 4 \beta Km^{2} \frac{\dot{r}^{2}}{r},\label{eq20}
\end{eqnarray}
the second term in parenthesis is very small compared to one. So, by neglecting it we have
\begin{eqnarray}
E=T+V&=&\frac{1}{2}m\dot{r}^{2}+ \frac{1}{2}m r^{2}\dot{\varphi}^{2}-\frac{K}{r}- 4 \beta K m^{2}\frac{\dot{r}^{2}}{r}\nonumber\\
&=&\frac{1}{2}(1-\frac{8 \beta m K}{r})m \dot{r}^{2}+\frac{1}{2}m r^{2} \dot{\varphi}^{2}- \frac{K}{r},\label{eq21}.
\end{eqnarray}
As can be seen, the contribution of the radial components of the kinetic energy is less than  ordinary classical mechanics by one.
\section{The orbits of planets }
Here we shall solve  equation (\ref{eq19}) to obtain the orbits of planets in the solar system. First, we define
\begin{equation}
u=\frac{1}{r},\label{eq22}
\end{equation}
then $ \dot{r}=-r^{2}\dot{\varphi}\frac{du}{d\varphi}=-\frac{L}{m}\frac{du}{d\varphi} $, $ \ddot{r}=-\frac{L^{2} u^{2}}{m^{2}}\frac{d^{2}u}{d\varphi^{2}}$ and the radial equation (\ref{eq19}) changes to
\begin{equation}
\frac{d^{2}u}{d\varphi^{2}}=-u+\frac{m}{L^{2}}K+ 4 K \beta m u^{2}+ 4 K \beta m (\frac{du}{d\varphi})^{2}.
\label{eq23}
\end{equation}
On the other hand, in the classical Kepler  problem we have
\begin{equation}
\frac{d^{2}u_{0}}{d\varphi^{2}}+u_{0}-\frac{1}{b}=0.\label{eq24}
\end{equation}
Solving this equation we obtain
\begin{equation}
u_{0}=\frac{1}{b}(1+ e\cos\varphi),\label{eq25}
\end{equation}
where $ e=\sqrt{1+\frac{2E L^{2}}{mK^{2}}} $ and $b=\frac{L^{2}}{mK}$. Now for solving  equation (\ref{eq23}), to first order in $\beta$, we suggest
\begin{equation}
u=u_{0}+u_{1},\label{eq26}
\end{equation}
where $ u_{1} $ must satisfy
\begin{equation}
\frac{d^{2}u_{1}}{d\varphi^{2}}+u_{1}=\frac{4 K\beta m}{b^{2}}[1+e^{2} +2 e\cos\varphi].,\label{eq27}
\end{equation}
By solving the equation above we  obtain
\begin{equation}
u_{1}=\frac{4 K\beta m}{b^{2}} [e \varphi \sin\varphi+1+e^{2}],\label{eq28}
\end{equation}
so we have
\begin{equation}
u= u_{0}+u_{1}= [\frac{1+e \cos(\varphi(1-\frac{\delta}{b}))}{b}]+\frac{\delta}{b^{2}}(1+ e^{2}),\label{eq29}
\end{equation}
that $ \delta= 4\beta K m $. Then we can obtain the perihelion shift per revolution as
\begin{equation}
\delta \varphi_{GUP}=\frac{2\pi\delta}{b}=2\pi(\frac{4\beta K m}{b}),\label{eq30}
\end{equation}
using the variables $ a=(\frac{b}{1-e^{2}}) $ and $ K=mGM_{\odot} $, where $ M_{\odot} $ is the  mass of the Sun, we have
\begin{equation}
\delta \varphi_{GUP}=2\pi( \frac{4 \beta m^{2} G M_{\odot}}{a(1-e^{2})}).\label{eq31}
\end{equation}
In the equation above, $ \beta $ the same as $ G $ and $ M_{\odot} $ is constant. In the case of the Mercury planet, using the data $ a\approx 6\times 10^{10}(m)$, $ e=0.2$, $ m\approx3.3\times 10^{23} (kg) $, $ M_{\odot}\approx 2.0\times 10^{30}(kg) $, $ G\approx 7.0 \times 10^{-11} (\frac{m^{3}}{s^{2} kg})$ and $ \hbar =6.6 \times 10^{-34} (Js)$, we found $ \delta \varphi_{GUP} $ as
\begin{equation}
\delta \varphi_{GUP}\approx2\pi\beta (1.06 \times 10^{57})(\frac{m^{2}kg^{2}}{s^{2}}).\label{eq33}
\end{equation}
For consistency  with the observational results, the $\delta\varphi_{GUP}$ must be small. Then we conclude that the parameter $\beta $ must be very small and the planetary system is very sensible in this parameter. So, small changes in $\beta$ implies sensible changes at very large scales. In other words, there is a connection between the physics at small scales and the physics at large scales. Now, we consider the observed perihelion shift of Mercury as \cite{Pireaux, Romero}
\begin{equation}
\delta\varphi_{obs}\approx 2\pi(7.98734\pm 0.00037)\times 10^{-8}(\frac{rad}{rev}),\label{eq34}
\end{equation}
if it is assumed that $\delta\varphi_{GUP}=\delta\varphi_{obs}$, we can obtain
\begin{equation}
\beta= 7.5352  \times 10^{-66} (\frac{s^{2}}{kg^{2} m^{2}}),\label{eq35}
\end{equation}
and
\begin{equation}
l_{min}=\hbar \sqrt{\beta}= 178.2\times 10^{-68} (m)=0.11\times 10^{-30}l_{pl}.\label{eq36}
\end{equation}
Note that this limit is $31$ orders of magnitude below the Planck length. However, for a best comparison it will be necessary to study the shift to the perihelion in a curved space. As we know,
in the case of General Relativity with the Schwarzschild-de Sitter
metric, the shift to the perihelion is \cite{Vatican}
\begin{equation}
\delta \phi_{GR,\Lambda}=6\pi\frac{G M_{\odot}}{c^{2}a(1-e^{2})}+\frac{\pi c^{2}\Lambda(1-e^{2})^{3} a^{3}}{GM_{\odot}},\label{eq32}
\end{equation}
so, for the perihelion shift of Mercury
\begin{equation}
\delta\varphi_{GR,\Lambda}=2\pi[7.98734\times 10^{-8}+6.1\times 10^{27}\Lambda](\frac{rad}{rev}),\label{eq37}
\end{equation}
now, if $\Lambda$ is of the order of $ 10^{-41}$ \cite{Kagr}, the second term above is the of same order as the observational error. Since the effect of spacetime quantization for a composite macroscopic body is much weaker than one of the constituent particles \cite{Giova}, we should consider
the effect of GUP on the perihelion shift as a perturbation effect which is very small. Using this assumption, we can find a upper bound for $\beta$ as
\begin{equation}
|\delta\varphi_{GUP}|\leq|\delta\varphi_{GR,\Lambda}-\delta\varphi_{obs}|\approx2\pi\times 10^{29}\times |\Lambda|(\frac{rad}{rev}),\label{eq38}
\end{equation}
then we obtain
\begin{equation}
\beta\leq 10^{-28}\times|\Lambda|(\frac{s^{2}}{kg^{2} m^{2}}),\label{eq39}
\end{equation}
and
\begin{equation}
l_{min}=\hbar \sqrt{\beta}\leq 6.6\times 10^{-48}\times\sqrt{|\Lambda|}= 4\times 10^{-13} \times\sqrt{|\Lambda|}\times  l_{pl},\label{eq40}
\end{equation}
by defining
\begin{equation}
l_{\Lambda}=\frac{1}{\sqrt{\Lambda}},\hspace{2.5cm} l_{0}^{2}=4\times 10^{-13} \times l_{pl},\label{eq41}
\end{equation}
the equation (\ref{eq40}) can been written as
\begin{equation}
l_{min}l_{\Lambda}=l_{0}^{2}.\label{eq42}
\end{equation}
If $\Lambda\rightarrow 0$ then $l_{min}\rightarrow 0$ and vice versa. It means that the cosmological constant of the spacetime is related to the minimal observable length. Without a the minimal length, there is no cosmological constant and vice versa. It can been seen that the minimal length $l_{min}$ and the cosmological constant scale $l_{\Lambda}=\frac{1}{\sqrt{\Lambda}}$ could be dual of each other\cite{Chang}-\cite{AKempf}.

On the other hand, There is an extensive literature suggesting that the relation $\Lambda \sim H^{2}$ plays a fundamental role in cosmology (here $H$ is
the Hubble parameter) \cite{Berman}. This relation has been obtained in a number of empirical models \cite{Chakra, Kall}.The dependence $\Lambda \sim H^{2}$ explains the current observations successfully and provides a more accurate age for the universe. According to this relation, the cosmological constant
varies with cosmic time and therefore the minimal length of spacetime decays with time. This matter could be very important in quantum gravity scenarios such as string theory.

\section{The effect of GUP on the stability of orbits}
For studying the stability of the planet circular orbits, we start from equation (\ref{eq16}). Using the general form of a central force $ F=-\frac{K}{r^{n}}$ we have
\begin{equation}
m(\ddot{r}-r \dot{\varphi}^{2})= - \frac{K}{r^{n}}- 4 \beta m^{2} ( \dot{r}^{2}+r^{2} \dot{\varphi}^{2} )\frac{K}{r^{n}},
\label{eq43}
\end{equation}
then, by considering $ L=mr^{2} \dot{\varphi} $, the effective potential can be written to first order in $\beta$ as
\begin{equation}
V_{eff}(r)=-\frac{1}{(n-1)}\frac{K}{r^{n-1}}+\frac{L^{2}}{2mr^{2}}- 4 \beta K\frac{L^{2}}{(n+1)r^{n+1}},\label{eq45}
\end{equation}
where $m$ is the mass of the planet and in a planet orbit $\dot{r}=0$. Now, the conditions of a stable circular orbit are
\begin{equation}
\frac{\partial V_{eff}}{\partial r}(r=r_{0})=0,\hspace{1.5cm} and \hspace{1.5cm}\frac{\partial^{2} V_{eff}}{\partial r^{2}}(r=r_{0}) > 0,\label{eq46}
\end{equation}
where $ r_{0}$ is the radius of orbit. Therefore, we should have
\begin{eqnarray}
\frac{\partial V_{eff}}{\partial r}=\frac{K}{r_{0}^{n}}-\frac{L^{2}}{m r_{0}^{3}}+{\beta K}\frac{L^{2}}{r_{0}^{n+2}}=0,\label{eq47}
\label{-6}
\end{eqnarray}
 and
\begin{equation}
\frac{\partial^{2} V_{eff}}{\partial r^{2}}(r=r_{0})=-\frac{nK}{r_{0}^{n+1}}+\frac{3 L^{2}}{m r_{0}^{4}}-\beta K\frac{(n+2)L^{2}}{r_{0}^{n+3}} > 0,\label{eq48}
\end{equation}
from the equation (\ref{eq47}) we have
\begin{equation}
\frac{L^{2}}{m r_{0}^{4}}=\frac{K}{r_{0}^{n+1}}+ \beta K \frac{L^{2}}{r_{0}^{n+3}},\label{eq49}
\end{equation}
using this result, the second condition for stability of circular orbits in central force can be written as
\begin{equation}
(3-n)K r_{0}^{2}- 4 (n-1)\beta K L^{2}>0,\label{eq50}
\end{equation}
this condition for stability differs from the one in general classical mechanics. When $\beta =0$ this results in $n<3$ which is a known condition in general classical mechanics. However, when $ \beta \neq 0 $, the condition for stability of circular orbits depends on the angular momentum. Since $ \beta $ is too small, the second term in equation (\ref{eq49}) is effective when the angular momentum of the planet is very large. Also, $ n $ could have non-integer values.

\section{Conclusion}
In this paper, we have investigate the effects of GUP on the Kepler problem. In other words we found the equation of motion  and the perihelion shift of the planets in this framework. Also, by comparing the value of the perihelion shift of Mercury in Schwarzschild-de Sitter spacetime with the resulted value of GUP, we showed that the minimal length $l_{min}$ and the cosmological constant scale $l_{\Lambda}$ could be dual of each other. Since the cosmological constant could depend on the
Hubble parameter, we concluded that the minimal length of the spacetime varies with cosmic time. In the end, we studied the stability of circular orbits of planets  and showed that the condition for stability is different from general classical mechanics, if the angular momentum of the planet is very large.
\section{Acknowledgment}
We wish to thank H. R. Sepangi for his careful reading of the article and for his constructive comments.


\end{document}